# Point defects and primary radiation damage in Co-free Fe–Ni–Cr–Mn high-entropy alloys from molecular dynamics

Ayobami Daramola[1,2,*], Giovanni Bonny[3], Gilles Adjanor[4], Christophe Domain[4], Ghiath Monnet[4], Anna Fraczkiewicz[5], Duc Nguyen-Manh[1,6,*]

[1] United Kingdom Atomic Energy Authority, Culham Campus, Abingdon, OX14 3DB, United Kingdom

[2] Centre for Science at Extreme Conditions and School of Physics and Astronomy, University of Edinburgh, Edinburgh, United Kingdom

[3] Nuclear Materials Science Institute, SCK CEN, Boeretang 200, B-2400, Mol, Belgium

[4] EDF Recherche et Développement, Département Matériaux et Mécanique des Composants, Electricité de France, Les Renardières, F-77250, Moret sur Loing, France

[5] CNRS, UMR 5307 LGF, Centre SMS, MINES Saint-Étienne, Université de Lyon, 42023, Saint-Étienne, France

[6] Department of Materials, University of Oxford, OX1 3PH, Oxford, United Kingdom

[*] Corresponding authors. E-mail: ayobami.daramola@ukaea.uk, [Duc email to be added]

## Abstract

Co-free Fe–Ni–Cr–Mn high-entropy alloys (HEAs) are candidates for nuclear structural applications. We used molecular dynamics with the EAM-21 potential, stiffened at short range, to study five HEAs that have been irradiated experimentally: $Fe_{37}Ni_{34}Cr_{16}Mn_{13}$ (ES1), $Fe_{35}Ni_{35}Cr_{15}Mn_{15}$ (P35), $Fe_{46}Ni_{22}Cr_{15}Mn_{17}$ (Y3), $Fe_{27}Ni_{28}Cr_{18}Mn_{27}$ (Kumar) and $Fe_{30}Ni_{30}Cr_{10}Mn_{30}$ (C30). They were compared with pure Ni, $Fe_{60}Ni_{40}$, the $Fe_{70}Ni_{10}Cr_{20}$ model alloy of 316L, and $Fe_{50}Ni_{34}Cr_{16}$, which is ES1 with Mn replaced by Fe. We calculated stacking fault energies, threshold displacement energies, point-defect formation and migration energies, and 1015 displacement cascades from 5 to 40 keV at 300 and 823 K. The stacking fault energy follows the Ni content: 19–44 mJ m$^{-2}$ in the alloys and 113 mJ m$^{-2}$ in Ni. Compared with the Mn-free alloys, the HEAs have lower threshold displacement energies (48–56 against 64–74 eV) and lower dumbbell formation energies, and Mn is the element most transported by interstitials. All the concentrated alloys leave about the same number of Frenkel pairs, 56–72 at 40 keV, half the number in Ni. At 40 keV, 54–58% of the interstitials are in clusters in the HEAs, against 67–71% in $Fe_{60}Ni_{40}$, $Fe_{70}Ni_{10}Cr_{20}$ and $Fe_{50}Ni_{34}Cr_{16}$ and 82% in Ni. At the same cluster size, clusters of 4–12 interstitials are faulted about half as often in the HEAs, and this does not follow the stacking fault energy. The surviving interstitials are enriched in Mn in the HEAs and in Ni in the Mn-free alloys, and depleted in Cr. The results support direct loop formation in cascades in $Fe_{60}Ni_{40}$ and $Fe_{70}Ni_{10}Cr_{20}$ and less of it in ES1.

# 1 Introduction

Austenitic Fe–Cr–Ni steels are used for core internals in light-water reactors and are candidates for advanced fission and fusion systems. Under irradiation they harden, lose ductility and swell, and these effects limit their life at high dose and temperature [1]. Concentrated fcc solid solutions with several principal elements have been proposed as an alternative, because some of them show less void swelling and smaller dislocation loops than conventional alloys [2–6]. The equiatomic CoCrFeMnNi (Cantor) alloy is the best known of these [7], but Co activates to $^{60}$Co under neutron irradiation. For nuclear use, Co-free Cr–Fe–Mn–Ni compositions have therefore been designed and tested [8–19]. The composition range in which these alloys stay single-phase fcc, and their tendency to order, have been mapped with first-principles cluster expansion and Monte Carlo simulations [20,21].

These Co-free alloys have now been irradiated under a range of conditions. Kumar et al. irradiated $Cr_{18}Fe_{27}Mn_{27}Ni_{28}$ with Ni ions between room temperature and 700 °C up to 10 dpa [8]. They found a high density of small dislocation loops, no voids, and radiation-induced segregation (RIS) at grain boundaries much weaker than in Fe–Cr–Ni alloys. Gao et al. irradiated $Cr_{15}Fe_{46}Mn_{17}Ni_{22}$ (alloy Y3) and $Fe_{37}Ni_{34}Cr_{16}Mn_{13}$ (alloy ES1) in situ with 2 MeV Fe ions from 95 to 823 K [9]. The loop density stayed near $5\times10^{22}$ m$^{-3}$ up to 623 K and fell to about $1\times10^{21}$ m$^{-3}$ at 823 K. At 823 K the loops were enriched in Ni and depleted in Cr, Fe and Mn. Parkin et al. compared $Cr_{18}Fe_{27}Mn_{27}Ni_{28}$ and $Cr_{15}Fe_{35}Mn_{15}Ni_{35}$ under dual-beam Kr and He irradiation at 500 and 600 °C [10]. Both swelled less than Ni and $Fe_{56}Ni_{44}$. The alloy with less Mn formed more faulted loops, and its measured stacking fault energy (SFE) was lower: 9.2±3.4 against 20.2±6.7 mJ m$^{-2}$. Curtis et al. neutron-irradiated $Cr_{10}Fe_{30}Mn_{30}Ni_{30}$ to 6.61 dpa at 395 and 579 °C [11]. They observed irradiation-enhanced local chemical order, no voids, and a faulted-loop density that decreased with dose.

Two recent in situ studies at 550 °C used $Fe_{37}Ni_{34}Cr_{16}Mn_{13}$ (alloy ES1). Dartois et al. compared it with $Fe_{60}Ni_{40}$ and 316L under the same conditions [12]. The quaternary alloy had larger loops and a higher growth rate, mainly through loop coarsening. Between 0.15 and 0.44 dpa no new loops formed in ES1, whereas $Fe_{60}Ni_{40}$ kept producing small loops. Dartois et al. took this as a sign that loops form directly in cascades in $Fe_{60}Ni_{40}$ but not in ES1. Ma et al. related the shape of Frank loops in Y3 and ES1 to their interstitial or vacancy character [13]. Mn content and SFE come up in all of these papers as possible reasons for the differences between alloys which was earlier hypothesized by Daramola et al. with atomistic modelling [22,23] . Ni content also changes between the alloys being compared, so the effect of Mn has not been separated.

Molecular dynamics (MD) has been used to look for the origin of radiation tolerance in concentrated alloys. Lin et al. found fewer surviving Frenkel pairs in NiCoCrFe than in Ni from 10 to 50 keV [24]. In their simulations, large interstitial clusters appeared from 20 keV in Ni but only from 30 keV in NiCoCrFe. Deluigi et al. compared an equiatomic FeNiCrCoCu alloy with an "average atom" material built from the same potential [25]. Frenkel-pair production and cluster sizes were the same within the statistics, and they concluded that any resistance must come from later defect evolution. Nair et al. analysed about 15000 cascades in Ni and five Fe–Ni–Cr compositions with several potentials [26]. Ni produced more defects than the alloys, the SFE given by the potential affected the size of the largest clusters, and potentials that are more repulsive at short range gave less cascade fragmentation and fewer defects. Hou et al. found that in $Fe_{46}Ni_{32}Cr_{22}$ the interstitials gather into large clusters, which become dislocation loops as the PKA energy increases [27]. Other studies examined local chemical order in CrFeNi [28], grain boundaries in FeCoCrNi [29], interstitial C and N in NiCoFeCrMn [30], cascade overlap in alloy 800H [31] and Fe content in FeCrNiAl [32]. Wei et al. showed for W–Ta and W–Mo that the short-range repulsion of each element pair controls the threshold displacement energy (TDE) and the number of Frenkel pairs [33]. Point defects in these alloys have also been studied by density functional theory (DFT). For $Cr_{18}Fe_{27}Mn_{27}Ni_{28}$, Li et al. reported a mean vacancy formation energy of 1.96 eV and vacancy migration barriers between 0.55 and 1.68 eV [34]. Comparable data exist for the Fe–10Ni–20Cr model alloy [35], CoCrFeMnNi [36,37] and dilute Ni–X alloys [38].

The MD studies above use model compositions, and the experiments use specific Co-free alloys. To our knowledge, no study has calculated primary damage and point-defect properties for the compositions that were actually irradiated, with one interatomic potential, and compared them with the measured trends. In this work we do this for five Co-free Fe–Ni–Cr–Mn high-entropy alloys (HEAs), ES1, P35, Y3, Kumar and C30, and for the Fe–10Ni–20Cr model alloy of 316L. We add pure Ni, $Fe_{60}Ni_{40}$, and a Mn-free version of ES1 in which Mn is replaced by Fe. We use the EAM-21 potential for Cr–Fe–Mn–Ni [39] with a stiffened short-range part. For each alloy we calculate the SFE, the TDE of each element, vacancy and interstitial formation energies, vacancy migration barriers and interstitial diffusion. We then simulate displacement cascades from 5 to 40 keV at 300 and 823 K. The cascades address three questions raised by the experiments. (i) Do interstitial clusters form directly in cascades in $Fe_{60}Ni_{40}$ and Fe–10Ni–20Cr but not in ES1 [12]? (ii) Does the fraction of faulted interstitial clusters follow the SFE [10]? (iii) Are the defects left by a cascade already enriched or depleted in particular elements before any diffusion, in the direction of the measured RIS [8,9]?

# 2 Methods

## 2.1 Alloys

The nine compositions are listed in Table 1. Five are Co-free Fe–Ni–Cr–Mn HEAs that have been irradiated experimentally: ES1, P35, Y3, Kumar and C30. They are named after the studies they come from; the alloy of Kumar et al. was also studied by Parkin et al. [8,10]. In the text these five are called the HEAs or the Mn-containing alloys [8–13]. $Fe_{70}Ni_{10}Cr_{20}$ is the Fe–10Ni–20Cr model alloy used to represent 316L [12,35,40]. $Fe_{60}Ni_{40}$ is the binary studied by Dartois et al. [12], and pure Ni is a reference. ES1noMn has the Ni and Cr content of ES1, with its 13 at.% Mn replaced by Fe. Comparing ES1 with ES1noMn gives the effect of Mn at fixed Ni and Cr. In all cells the elements were placed at random on the fcc sites with the exact atomic counts; no short-range order was imposed. All simulations were run with LAMMPS [41].

***Table 1.*** *Alloys (compositions in at.%), the experiments they come from, and properties calculated with the potential used here: lattice parameter at 0, 300 and 823 K (Å), fraction of fcc atoms from common neighbour analysis [42] after 20 ps at 823 K, and cohesive energy (eV).*

| Alloy | Fe | Ni | Cr | Mn | Ref. | $a_{0\,K}$ | $a_{300\,K}$ | $a_{823\,K}$ | $fcc_{823\,K}$ | $E_{coh}$ |
|---|---|---|---|---|---|---|---|---|---|---|
| Ni | 0 | 100 | 0 | 0 | – | 3.5193 | 3.5197 | 3.5315 | 0.994 | −4.4497 |
| Fe60Ni40 | 60 | 40 | 0 | 0 | [12] | 3.4873 | 3.4959 | 3.5172 | 0.996 | −4.4220 |
| Fe70Ni10Cr20 | 70 | 10 | 20 | 0 | [12,35] | 3.5160 | 3.5252 | 3.5514 | 0.985 | −4.1694 |
| ES1noMn | 50 | 34 | 16 | 0 | this work | 3.5031 | 3.5128 | 3.5373 | 0.991 | −4.3115 |
| ES1 | 37 | 34 | 16 | 13 | [12,13] | 3.5033 | 3.5129 | 3.5356 | 0.990 | −4.1258 |
| P35 | 35 | 35 | 15 | 15 | [10] | 3.5026 | 3.5121 | 3.5347 | 0.990 | −4.1081 |

| Alloy | Fe | Ni | Cr | Mn | Ref. | $a_{0\,\mathrm{K}}$ | $a_{300\,\mathrm{K}}$ | $a_{823\,\mathrm{K}}$ | $\mathrm{fcc}_{823\,\mathrm{K}}$ | $E_{coh}$ |
|---|---|---|---|---|---|---|---|---|---|---|
| Y3 | 46 | 22 | 15 | 17 | [9,13,15, 16] | 3.5036 | 3.5132 | 3.5362 | 0.990 | −4.0284 |
| Kumar | 27 | 28 | 18 | 27 | [8,10] | 3.5070 | 3.5171 | 3.5396 | 0.987 | −3.8971 |
| C30 | 30 | 30 | 10 | 30 | [11] | 3.4984 | 3.5079 | 3.5287 | 0.991 | −3.9101 |

## 2.2 Interatomic potential

We used the EAM-21 embedded-atom potential for Cr–Fe–Mn–Ni [39], which was fitted for the plasticity of these alloys [23]. It has also been used to study the unfaulting of interstitial Frank loops and their interaction with dislocations in Y3 and $Fe_{70}Ni_{10}Cr_{20}$ [22]. For cascade simulations its short-range part was stiffened. Each pair function follows the Ziegler–Biersack–Littmark (ZBL) universal potential [43] to within a few per cent below about 1.0 Å and joins the EAM-21 pair function by 1.97–2.19 Å, depending on the pair. The electron density functions are held constant at short range and equal the EAM-21 functions beyond 2.09 Å. This keeps the host density in a close collision inside the range of the embedding functions. For the Fe–Mn, Ni–Mn and Cr–Mn pairs, the stiffened form below 1.0 Å was joined to the published EAM-21 functions with a quintic ($C^2$-continuous) switching function between 1.0 and 2.0 Å. All three blended pairs decrease monotonically from 0.05 to 2.3 Å. The nearest-neighbour distance in these alloys is about 2.48 Å, so all changes lie well inside the first shell. The stiffened potential reproduces the lattice parameter and cohesive energy of the original EAM-21 for all nine alloys to within $10^{-4}$ Å and $10^{-4}$ eV/atom. The pair and density functions are compared with EAM-21 and ZBL in Fig. S1.

All cells in this work were built with the lattice parameter of the potential at the relevant temperature (Table 1).

## 2.3 Stacking fault energy

The intrinsic SFE was calculated in slabs oriented with $x=[11\bar{2}]$, $y=[111]$ and $z=[1\bar{1}0]$. Each slab contained 19 (111) planes and 6080 atoms, with periodic boundaries along $x$ and $z$ and free surfaces normal to $y$. The fault area was about 34×50 Å$^2$ (≈1700 Å$^2$). After relaxation of the perfect slab, the upper nine planes were displaced rigidly by one Shockley partial, $b_p=a/\sqrt{6}$, along $[11\bar{2}]$. The slab was then relaxed with atomic motion allowed only along [111]. Both senses of the shift were applied, and the lower energy corresponds to the intrinsic fault. Common neighbour analysis confirmed that this configuration contains exactly two hcp (111) planes, while the other sense contains none. The SFE is

$$\gamma_{isf} = \frac{E_{SF} - E_0}{A},$$

where $E_{SF}$ and $E_0$ are the energies of the faulted and perfect slabs and $A$ is the fault area. The surface energy cancels in the difference. For each alloy, 40 slabs with different random arrangements of the elements were used. We report the mean and its standard error. The spread between slabs depends on the fault area and is given for this area only.

## 2.4 Threshold displacement energy

Threshold displacement energies were obtained in 6912-atom cells (12×12×12 cubic unit cells) at the 300 K lattice parameter of each alloy, equilibrated for 10 ps at 300 K. We used 100 random recoil directions, uniformly distributed on the unit sphere. The same set of directions was used for every alloy and element, so that alloys can be compared direction by direction. For each element and direction, the primary knock-on atom (PKA) was a randomly chosen atom of that element within 6 Å of the cell centre. Its kinetic energy was increased from 10 eV in steps of 2 eV. Each trial started from the same equilibrated state and was followed for 4 ps in the NVE ensemble with an adaptive time step (at most 0.02 Å of atomic displacement per step, $\Delta t \leq 1$ fs). The cell was then relaxed to 0 K. A stable Frenkel pair was taken to have formed when common neighbour analysis found at least 8 non-fcc atoms. An isolated vacancy alone makes its 12 neighbours non-fcc, while recombined or replacement-only events leave none. The TDE of a direction is the lowest energy at which a stable defect formed. Directions without a stable defect at 150 eV were recorded as above 150 eV. Because these directions make the mean ill-defined, we report medians and interquartile ranges. An effective TDE for each alloy, $E_d^{\mathrm{eff}}$, was taken as the median over all directions and elements, with each element weighted by its concentration. Electronic stopping was not included in the TDE runs.

## 2.5 Point-defect formation energies

Point-defect energies were calculated in 2048-atom cells (8×8×8 unit cells) at the 0 K lattice parameter, at constant volume. Formation energies in a concentrated alloy need the chemical potential $\mu_i$ of each element. Replacing one atom of element $h$ by element $s$ changes the energy of the relaxed cell by $\Delta E_{h\rightarrow s} \approx \mu_s - \mu_h$. For each alloy we calculated 60 such swaps for each target element, at random sites. The $\mu_i$ were then obtained by least squares from the swap energies, with the constraint

$$\sum_i x_i \, \mu_i = \frac{E_0}{N},$$

where $x_i$ is the concentration of element $i$ and $E_0$ is the energy of the perfect $N$-atom cell. The vacancy formation energy at a site occupied by element $i$, and the formation energy of a ⟨100⟩ dumbbell made by adding an atom of element $s$, are

$$E_v^f(i) = E_{N-1} - E_0 + \mu_i, E_I^f(s) = E_{N+1} - E_0 - \mu_s.$$

We calculated 60 vacancies per element and 60 dumbbells per added element. For a dumbbell, the host atom at a random site was moved by $-0.3a$ along a random ⟨100⟩ direction, the new atom was placed at $+0.3a$, and the cell was relaxed. Alloy averages were taken over elements, weighted by concentration.

## 2.6 Defect migration

Vacancy migration barriers were calculated with the climbing-image nudged elastic band (NEB) method [44] with seven replicas. In each alloy, 30 vacancy sites were chosen at random. For each site, one nearest neighbour of each element present was moved into the vacancy, giving 30–115 jumps per alloy. Both end states were relaxed before the NEB calculation, which was converged to a maximum force of 0.01 eV/Å on any replica. We report the forward barrier, measured from the state with the vacancy at the chosen site, and sort the jumps by the element that moves. Calculations that did not converge or gave barriers outside 0.05–4 eV were discarded.

Dumbbell interstitials in a random alloy can migrate along many paths, so their mobility was obtained from MD. One ⟨100⟩ dumbbell was inserted in a 2048-atom cell, and the cell was run for 4 ns in the NVT ensemble at 600, 750, 900 and 1050 K, with six independent runs per temperature. The cell size at each temperature was set from the lattice parameters in Table 1 by linear interpolation. We followed the summed squared displacement of all atoms from their starting positions, $R^2(t) = \sum_j \left|r_j(t) - r_j(0)\right)^2$. The interstitial diffusion coefficient was taken as $D$=(d$R^2$/d$t$)/6, fitted for $t$>0.2 ns, and an effective migration energy $E_m^I$ was obtained from an Arrhenius fit of $D(T)$. The same sum taken over the atoms of one element gives the share of interstitial transport carried by that element. Divided by its concentration, this gives a transport ratio. A value above one means interstitials move that element more than its share. Transport ratios were taken from the runs at 900 and 1050 K, where each run contains many jumps.

## 2.7 Displacement cascades

Cascades were simulated at PKA energies of 5, 10, 20 and 40 keV in all nine alloys at 300 K. ES1, ES1noMn, Y3 and $Fe_{70}Ni_{10}Cr_{20}$ were also run at 10 and 20 keV at 823 K, close to the temperature of the in situ experiments [9,12,13]. Cubic cells with periodic boundaries were used, with edges of 40, 56, 72 and 96 unit cells (about 140, 197, 253 and 337 Å; 0.26, 0.70, 1.49 and 3.54 million atoms) at 5, 10, 20 and 40 keV. These sizes were chosen from a pilot study of 24 cascades in ES1 (Fig. S2). In cells of these sizes the main cascade region stayed clear of the thermostat. Single channelled recoils still reached the cell faces in some runs, in all cell sizes tested, and occasionally left an isolated Frenkel pair there. Each cell was equilibrated for 10 ps at the target temperature in the NVT ensemble, at the lattice parameter of the potential for that temperature. During the cascade, atoms within 7 Å of the cell faces were coupled to a Langevin thermostat at the same temperature with a damping time of 0.1 ps. The rest of the cell was integrated in the NVE ensemble.

The PKA element was drawn according to the alloy composition and the recoil direction at random. To keep the cascade near the cell centre, the PKA was the atom of that element nearest to a point displaced from the centre by 15, 25 or 40 Å (at 10, 20 and 40 keV) opposite to the recoil direction. The time step was adaptive, with at most 0.02 Å of atomic displacement per step and $\Delta t \leq 0.5$ fs for the first 5 ps, and 0.05 Å and $\Delta t \leq 1$ fs afterwards. Each cascade was run for 30 ps. At 300 K we ran 20 cascades per alloy at 5 keV, 30 at 10 and 20 keV, and 15 at 40 keV; at 823 K we ran 20 per alloy at 10 and 20 keV, giving 1015 cascades in total.

Electronic stopping was applied as a friction force on atoms with kinetic energy above 10 eV. Stopping powers were taken from SRIM-2013 [45] for each ion (Fe, Ni, Cr, Mn) in each alloy. Per-element stopping cross sections were obtained by least squares from SRIM tables calculated for seven of the compositions with Bragg additivity; they reproduce these tables to within 0.03%. Tables for all nine alloys were then built at the atomic density given by the potential at 300 K. Below 100 eV the stopping was extended as $S \propto E^{1/2}$, which is the form the tabulated values follow at low energy.

Defects were identified by Wigner–Seitz analysis with the perfect lattice as reference. The lattice sites were stored when each cell was built. An atom less than 0.9 Å from its own site is inside its own Wigner–Seitz cell, since the distance from a site to the cell boundary is half the nearest-neighbour distance (1.24 Å). Only atoms more than 0.9 Å from their own site were therefore written out, every 0.1 ps up to 2 ps, every 0.5 ps up to 5 ps and every 2.5 ps up to 30 ps. Each written atom was assigned to its nearest lattice site; a vacancy is a site with no atom and an interstitial is an extra atom at a site. This gives the exact Wigner–Seitz count without writing whole configurations. The number of displaced atoms was also recorded every 50 steps to obtain its peak value and the duration of the thermal spike. For every cascade we recorded the number of surviving defects within 7 Å plus one lattice parameter of a cell face. Results are given for all cascades and, as a check, without the cascades that left defects in this region.

## 2.8 Analysis of the cascade damage

The damage energy $E_{\mathrm{dam}}$ of each PKA was obtained from the Robinson partition used in the NRT model [46], with the concentration-weighted atomic number and mass of the alloy. The NRT number of Frenkel pairs is $N_{\mathrm{NRT}}=0.8E_{\mathrm{dam}}/(2E_d)$, where we used $E_d=E_d^{\mathrm{eff}}$ of each alloy (Section 3.3). The survival efficiency $\xi=N_{\mathrm{FP}}/N_{\mathrm{NRT}}$ was compared with the arc-dpa form [47],

$$\xi(E_{dam}) = \frac{1-c}{(2\,E_d/0.8)^b} E_{dam}^b + c,$$

with the parameters given for Ni, $b=-1.007$ and $c=0.227$ [47]. No arc-dpa parameters exist for the alloys studied here.

Surviving vacancies and interstitials were grouped into clusters separately. Two defects belong to the same cluster if their lattice sites are within one lattice parameter $a$, which includes first and second neighbours. A defect in a cluster of two or more is counted as clustered. For each cascade we recorded the fraction of interstitials and of vacancies that are clustered and the size of the largest interstitial cluster.

To differentiate faulted from unfaulted interstitial clusters, the final configuration of every cascade at 10, 20 and 40 keV (835 cascades, including those at 823 K) was relaxed by conjugate gradient minimisation, and each atom was classified by polyhedral template matching (PTM) [48] with an RMSD cutoff of 0.1. Relaxation removes the thermal noise that otherwise produces scattered non-fcc atoms. An interstitial cluster of at least four interstitials was counted as faulted if at least four hcp atoms lay within 1.5$a$ of its members. A faulted loop contains an intrinsic or extrinsic stacking fault, which PTM shows as hcp atoms, while an unfaulted cluster or a perfect loop has none.

The dislocation content of the 40 keV cascades (135 cascades) was obtained with the dislocation extraction algorithm (DXA) [49] in OVITO 3.15 [50], applied to the final configurations at 300 K. The total line length was split by Burgers vector into Shockley ($1/6\langle 112\rangle$), stair-rod ($1/6\langle 110\rangle$), Frank ($1/3\langle 111\rangle$), Hirth ($1/3\langle 100\rangle$) and perfect ($1/2\langle 110\rangle$) segments. For the snapshots in Fig. 7, one 40 keV cascade was chosen per alloy. It is the cascade without defects near the faces whose largest interstitial cluster is closest to the mean of that alloy. The close-ups show the non-fcc atoms, from adaptive common neighbour analysis, within a sphere around the largest interstitial cluster; only the largest connected group of these atoms is drawn.

# 3 Results

## 3.1 Lattice parameter and phase stability

All nine alloys stayed fcc. After 20 ps the fraction of fcc atoms was 1.000 at 300 K and 0.985–0.996 at 823 K, where the small deficit comes from thermal vibrations in the common neighbour analysis (Table 1). $Fe_{70}Ni_{10}Cr_{20}$ also remained fcc. The phase stability of Cr–Fe–Mn–Ni alloys over the full composition range has been calculated by Fedorov et al. [21]. The lattice parameter at 300 K ranges from 3.496 Å ($Fe_{60}Ni_{40}$) to 3.525 Å ($Fe_{70}Ni_{10}Cr_{20}$). Replacing 13 at.% Fe by Mn hardly changes it (ES1noMn 3.5128 Å, ES1 3.5129 Å). Between 300 and 823 K the lattice parameter increases by 0.34% in Ni and by 0.61–0.74% in the alloys.

## 3.2 Stacking fault energy

***Table 2.*** *Intrinsic stacking fault energy $\gamma_{isf}$ in mJ m$^{-2}$: mean ± standard error over 40 slabs (fault area ≈1700 Å$^2$), with measured values.*

| Alloy | Ni (at.%) | Mn (at.%) | This work | Measured |
|---|---|---|---|---|
| Fe70Ni10Cr20 | 10 | 0 | 19.1±0.8 | |
| Y3 | 22 | 17 | 31.0±1.2 | |
| Kumar | 28 | 27 | 36.7±1.1 | 20.2±6.7 [10] |
| C30 | 30 | 30 | 39.1±1.3 | |
| ES1noMn | 34 | 0 | 36.3±1.0 | |
| ES1 | 34 | 13 | 41.1±1.1 | |
| P35 | 35 | 15 | 39.9±1.7 | 9.2±3.4 [10] |
| Fe60Ni40 | 40 | 0 | 44.3±0.9 | |
| Ni | 100 | 0 | 113.4 | 110–130 [51,52] |

The SFE is controlled mainly by the Ni content (Table 2, Fig. 1). It rises from 19.1 mJ m$^{-2}$ in $Fe_{70}Ni_{10}Cr_{20}$ (10% Ni) to 44.3 mJ m$^{-2}$ in $Fe_{60}Ni_{40}$ (40% Ni), and is 113.4 mJ m$^{-2}$ in Ni. At fixed Ni and Cr, 13 at.% Mn raises the SFE from 36.3±1.0 to 41.1±1.1 mJ m$^{-2}$, an increase of 4.8±1.5 mJ m$^{-2}$. Across the Mn-containing alloys the values do not follow the Mn content: Y3 (22% Ni, 17% Mn) has 31.0 mJ m$^{-2}$ and ES1 (34% Ni, 13% Mn) has 41.1 mJ m$^{-2}$.

Parkin et al. measured 9.2 mJ m$^{-2}$ for P35 and 20.2 mJ m$^{-2}$ for Kumar [10]; we obtain 39.9 and 36.7 mJ m$^{-2}$. The two alloys differ in Ni (35 and 28%) as well as in Mn (15 and 27%), and in our calculations the effect of Ni is the larger one. The standard deviation between slabs is 5–9 mJ m$^{-2}$ for a 1700 Å$^2$ fault.

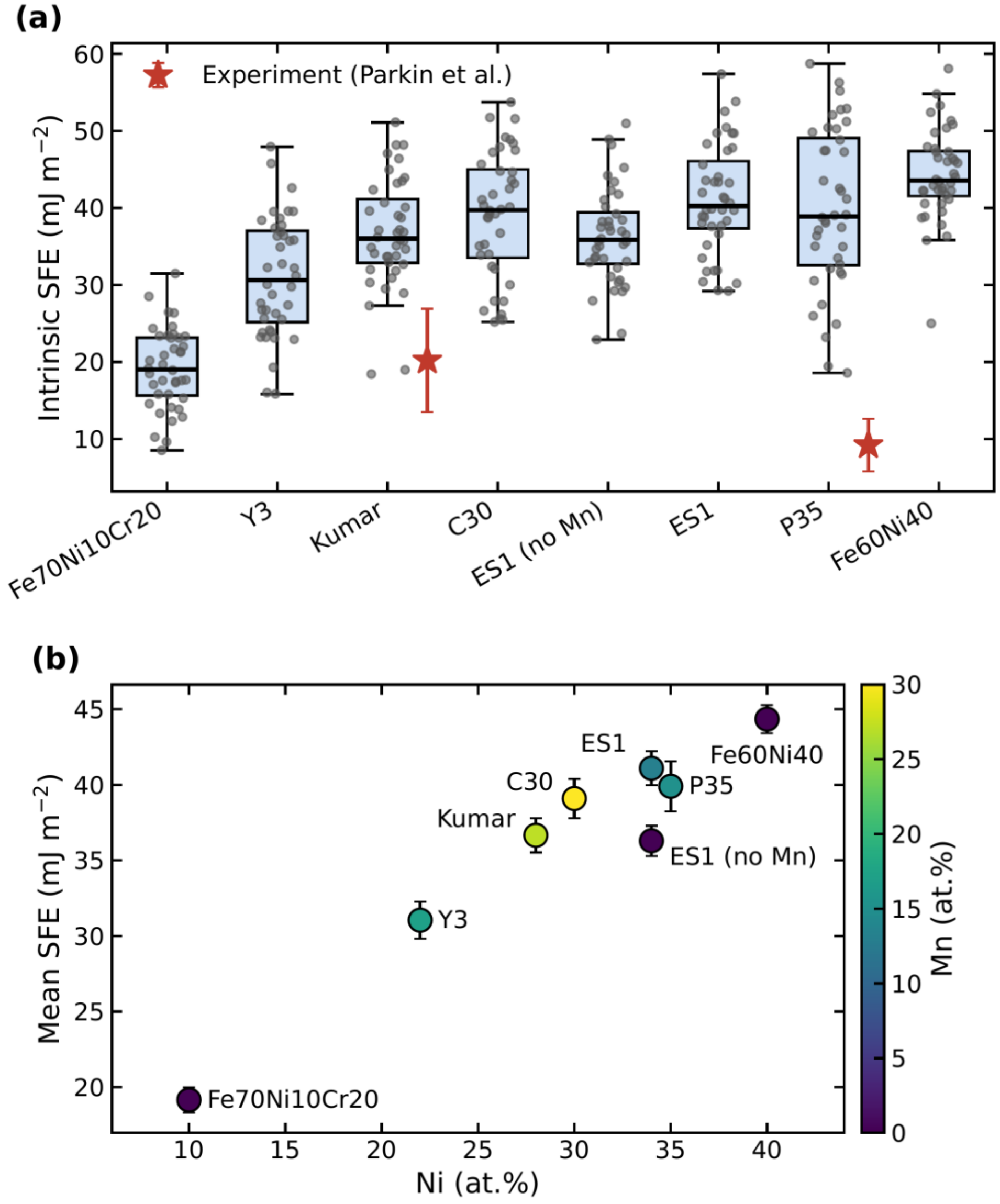

***Fig. 1.*** *Intrinsic stacking fault energy. (a) Distribution over 40 slabs for each alloy (box: quartiles; points: individual slabs), with the values measured by Parkin et al. [10] (stars). (b) Mean SFE against Ni content, coloured by Mn content; error bars are standard errors.*

## 3.3 Threshold displacement energy

***Table 3.*** *Threshold displacement energy: median over 100 directions for each recoiling element (minimum in brackets), percentage of directions above 150 eV, and the concentration-weighted alloy value $E_d^{eff}$. All energies in eV.*

| Alloy | Fe | Ni | Cr | Mn | >150 eV (%) | $E_d^{eff}$ |
|---|---|---|---|---|---|---|
| Ni | – | 54 (22) | – | – | 0 | 54 |
| Fe60Ni40 | 65 (34) | 61 (34) | – | – | 6–9 | 64 |
| Fe70Ni10Cr20 | 75 (38) | 70 (26) | 73 (34) | – | 6–10 | 74 |
| ES1noMn | 66 (34) | 61 (22) | 72 (30) | – | 2–9 | 66 |
| ES1 | 59 (28) | 51 (24) | 61 (26) | 53 (28) | 1–7 | 56 |
| P35 | 54 (28) | 50 (26) | 61 (30) | 52 (16) | 1–3 | 54 |
| Y3 | 55 (32) | 51 (28) | 60 (32) | 58 (22) | 0–8 | 56 |

| Alloy | Fe | Ni | Cr | Mn | >150 eV (%) | $E_d^{eff}$ |
|---|---|---|---|---|---|---|
| Kumar | 50 (22) | 48 (24) | 51 (28) | 49 (18) | 1–2 | 50 |
| C30 | 46 (24) | 47 (22) | 48 (28) | 50 (18) | 0–2 | 48 |

The TDE results are given in Table 3 and Fig. 2, and the full statistics for each element in Table S1. In Ni the median is 54 eV and the lowest direction gives 22 eV, close to the measured minimum threshold of about 23 eV [53]. Since the same directions were used for every alloy, the alloys can be compared directly.

$E_d^{eff}$ is 64–74 eV in the Mn-free alloys and 48–56 eV in the Mn-containing alloys. $Fe_{70}Ni_{10}Cr_{20}$ has the highest value (74 eV) and C30 the lowest (48 eV). Within an alloy the elements differ by less than about 10 eV. Ni recoils have the lowest median in most alloys and Cr recoils the highest. Because the same directions were used in ES1 and ES1noMn, the two alloys can be compared direction by direction. The median of the paired differences (ES1 minus ES1noMn, with 95% bootstrap intervals) is 0 eV [−6, +2] for Fe recoils, −6 eV [−12, 0] for Ni and −6 eV [−10, 0] for Cr. At fixed Ni and Cr, 13% Mn therefore lowers the TDE of Ni and Cr recoils by about 6 eV and does not change it for Fe. Most of the difference between the Mn-free and the Mn-containing alloys in Table 3 comes from the Fe and Ni contents, which change together with Mn across this set; the most Fe-rich alloys have the highest TDE.

In the Fe-rich Mn-free alloys, up to 10% of directions did not produce a stable defect at 150 eV. These directions enter the medians as values above 150 eV.

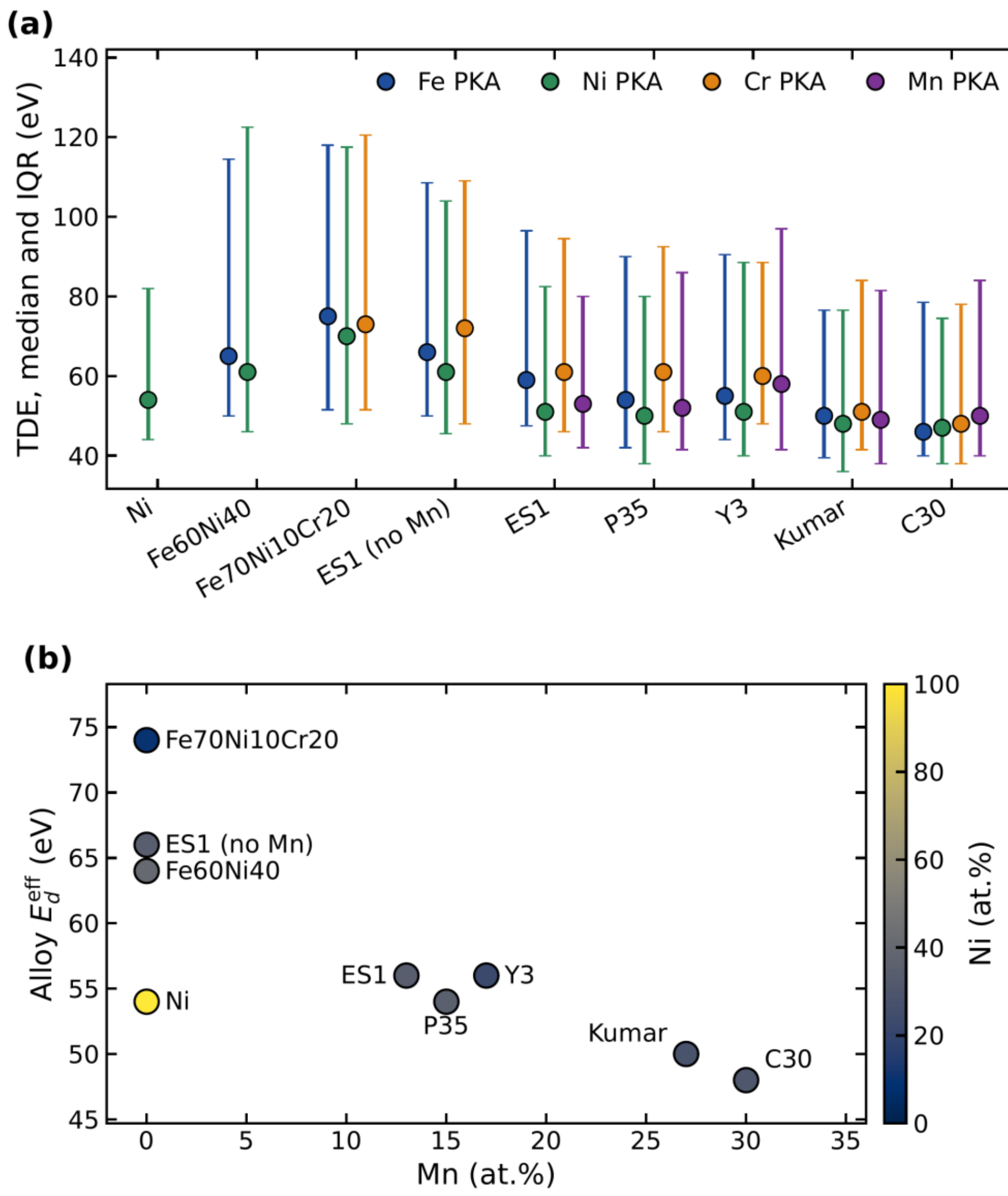

***Fig. 2.*** *Threshold displacement energy. (a) Median and interquartile range for each recoiling element. (b) Concentration-weighted alloy value $E_d^{eff}$ against Mn content, coloured by Ni content.*

## 3.4 Point-defect formation energies

***Table 4.*** *Vacancy formation energy by the element removed, and ⟨100⟩ dumbbell formation energy by the element added, with concentration-weighted alloy averages (eV). Standard deviations over 60 configurations are 0.05–0.16 eV for vacancies and 0.2–0.55 eV for dumbbells.*

| | $E_v^f$ | | | | | $E_I^f$ | | | | |
|---|---|---|---|---|---|---|---|---|---|---|
| Alloy | Fe | Ni | Cr | Mn | alloy | Fe | Ni | Cr | Mn | alloy |
| Ni | – | 1.48 | – | – | 1.48 | – | 4.09 | – | – | 4.09 |
| Fe60Ni40 | 1.93 | 1.93 | – | – | 1.93 | 5.12 | 4.86 | – | – | 5.02 |
| Fe70Ni10Cr20 | 2.00 | 2.03 | 2.04 | – | 2.01 | 4.26 | 4.17 | 4.49 | – | 4.29 |
| ES1noMn | 1.95 | 1.95 | 1.98 | – | 1.95 | 4.84 | 4.57 | 4.89 | – | 4.76 |
| ES1 | 1.92 | 1.92 | 1.91 | 1.91 | 1.92 | 4.39 | 4.37 | 4.53 | 4.07 | 4.36 |
| P35 | 1.91 | 1.89 | 1.89 | 1.90 | 1.90 | 4.30 | 4.37 | 4.54 | 4.00 | 4.32 |
| Y3 | 1.96 | 1.95 | 1.98 | 1.97 | 1.96 | 4.31 | 4.31 | 4.33 | 3.83 | 4.23 |
| Kumar | 1.91 | 1.96 | 1.93 | 1.91 | 1.93 | 4.17 | 4.02 | 4.13 | 3.75 | 4.01 |
| C30 | 1.91 | 1.87 | 1.89 | 1.89 | 1.89 | 4.01 | 4.11 | 4.23 | 3.83 | 4.01 |

Vacancy formation energies are given in Table 4 and Fig. 3(a); the chemical potentials and the standard deviations are in Table S2. In Ni the value is 1.48 eV, close to the DFT value of 1.40 eV [38]. In the alloys it lies between 1.89 and 2.01 eV and depends little on the element removed (by at most 0.05 eV within an alloy). For $Cr_{18}Fe_{27}Mn_{27}Ni_{28}$ we obtain 1.93 eV, against a DFT mean of 1.96 eV [34]. For $Fe_{70}Ni_{10}Cr_{20}$ we obtain 2.01 eV, within the DFT range of 1.76–2.19 eV (mean 1.96 eV) for Fe–10Ni–20Cr [35]. DFT values for CoCrFeMnNi are similar, 1.88 eV [37] and about 2.0 eV [36].

The dumbbell formation energy is 4.09 eV in Ni and 4.01–5.02 eV in the alloys. It is lower in the Mn-containing alloys (4.01–4.36 eV) than in the Mn-free alloys (4.29–5.02 eV). In every Mn-containing alloy, dumbbells formed by adding Mn have the lowest energy (3.75–4.07 eV) and those formed by adding Cr have the highest (4.13–4.54 eV). Mn is therefore the preferred element in interstitial sites and Cr the least favoured one.

## 3.5 Vacancy migration

***Table 5.*** *Vacancy migration barriers (mean over jumps, eV) by the element that jumps into the vacancy. Number of jumps in brackets; standard deviations are 0.08–0.19 eV.*

| Alloy | Fe | Ni | Cr | Mn |
|---|---|---|---|---|
| Ni | – | 1.17 (30) | – | – |
| Fe60Ni40 | 0.99 (30) | 1.37 (30) | – | – |
| Fe70Ni10Cr20 | 0.85 (30) | 1.23 (23) | 1.11 (27) | – |
| ES1noMn | 0.98 (30) | 1.26 (30) | 1.18 (26) | – |
| ES1 | 1.03 (29) | 1.31 (30) | 1.18 (22) | 1.11 (22) |
| P35 | 1.08 (30) | 1.23 (30) | 1.14 (27) | 1.08 (25) |
| Y3 | 1.04 (30) | 1.23 (28) | 1.15 (26) | 1.09 (28) |
| Kumar | 1.10 (30) | 1.24 (29) | 1.12 (28) | 1.01 (28) |
| C30 | 1.11 (30) | 1.24 (30) | 1.13 (22) | 0.97 (29) |

The vacancy migration barriers are given in Table 5 and Fig. 3(b). In every alloy, Ni jumps have the highest barriers (1.23–1.37 eV, against 1.17 eV in pure Ni). The lowest barriers belong to Fe jumps in the Mn-free alloys, ES1 and Y3 (0.85–1.04 eV), and to Mn jumps in Kumar and C30 (1.01 and 0.97 eV). The Fe barrier increases with Mn content, from 0.85 eV in $Fe_{70}Ni_{10}Cr_{20}$ to 1.10–1.11 eV in Kumar and C30. The high Ni barrier agrees with DFT for CoCrFeMnNi, where Ni and Co have the highest barriers [36,37], and with the slow Ni–vacancy exchange invoked by Dartois et al. [12]. The standard deviation of the barriers within an alloy is 0.08–0.19 eV.

## 3.6 Interstitial diffusion

***Table 6.*** *Interstitial migration from MD: effective migration energy $E_m^I$ and prefactor $D_0$ from an Arrhenius fit between 600 and 1050 K, and transport ratio of each element (share of the summed squared displacement divided by concentration) from the runs at 900 and 1050 K.*

| Alloy | $E_m^I$ (eV) | $D_0$ ($m^2 s^{-1}$) | Fe | Ni | Cr | Mn |
|---|---|---|---|---|---|---|
| Ni | 0.31 | $2.2\times10^{-7}$ | – | 1.00 | – | – |
| Fe60Ni40 | 0.40 | $2.9\times10^{-7}$ | 0.30 | 2.06 | – | – |
| Fe70Ni10Cr20 | 0.34 | $2.9\times10^{-7}$ | 1.11 | 1.96 | 0.13 | – |
| ES1noMn | 0.54 | $8.5\times10^{-7}$ | 0.37 | 2.32 | 0.14 | – |
| ES1 | 0.40 | $2.1\times10^{-7}$ | 0.43 | 0.85 | 0.06 | 4.17 |
| P35 | 0.42 | $3.2\times10^{-7}$ | 0.45 | 0.79 | 0.05 | 3.73 |
| Y3 | 0.53 | $8.5\times10^{-7}$ | 0.52 | 0.59 | 0.03 | 3.68 |
| Kumar | 0.38 | $2.2\times10^{-7}$ | 0.42 | 0.68 | 0.06 | 2.54 |
| C30 | 0.38 | $2.6\times10^{-7}$ | 0.42 | 0.53 | 0.04 | 2.38 |

No dumbbell jump occurred at 300 K in any alloy during 2 ns test runs, so the interstitial mobility was measured between 600 and 1050 K (Table 6, Fig. 3c). The effective migration energy is 0.31 eV in Ni. In the alloys $E_m^I$ is 0.34–0.42 eV, except in ES1noMn and Y3, where it is 0.53–0.54 eV. Interstitial migration is therefore slower and more strongly activated in the concentrated alloys than in Ni. At fixed Ni and Cr, Mn lowers the migration energy from 0.54 eV (ES1noMn) to 0.40 eV (ES1).

The transport ratios show which elements the interstitials move (Table 6, Fig. 3d). Cr is hardly moved by interstitials in any alloy (ratio 0.03–0.14). In the Mn-free alloys Ni is moved about twice as often as its concentration would give (1.96–2.32). In the Mn-containing alloys Mn takes this role (2.38–4.17) and the Ni ratio falls below one (0.53–0.85). Fe is under-represented except in $Fe_{70}Ni_{10}Cr_{20}$. This follows the dumbbell formation energies (Table 4): interstitials spend most time in Mn-containing dumbbells, or Ni-containing ones when there is no Mn, and avoid Cr. An interstitial flux to sinks would then carry Mn (or Ni) towards them and leave Cr behind. The Cr depletion agrees with the Cr depletion measured at loops [9].

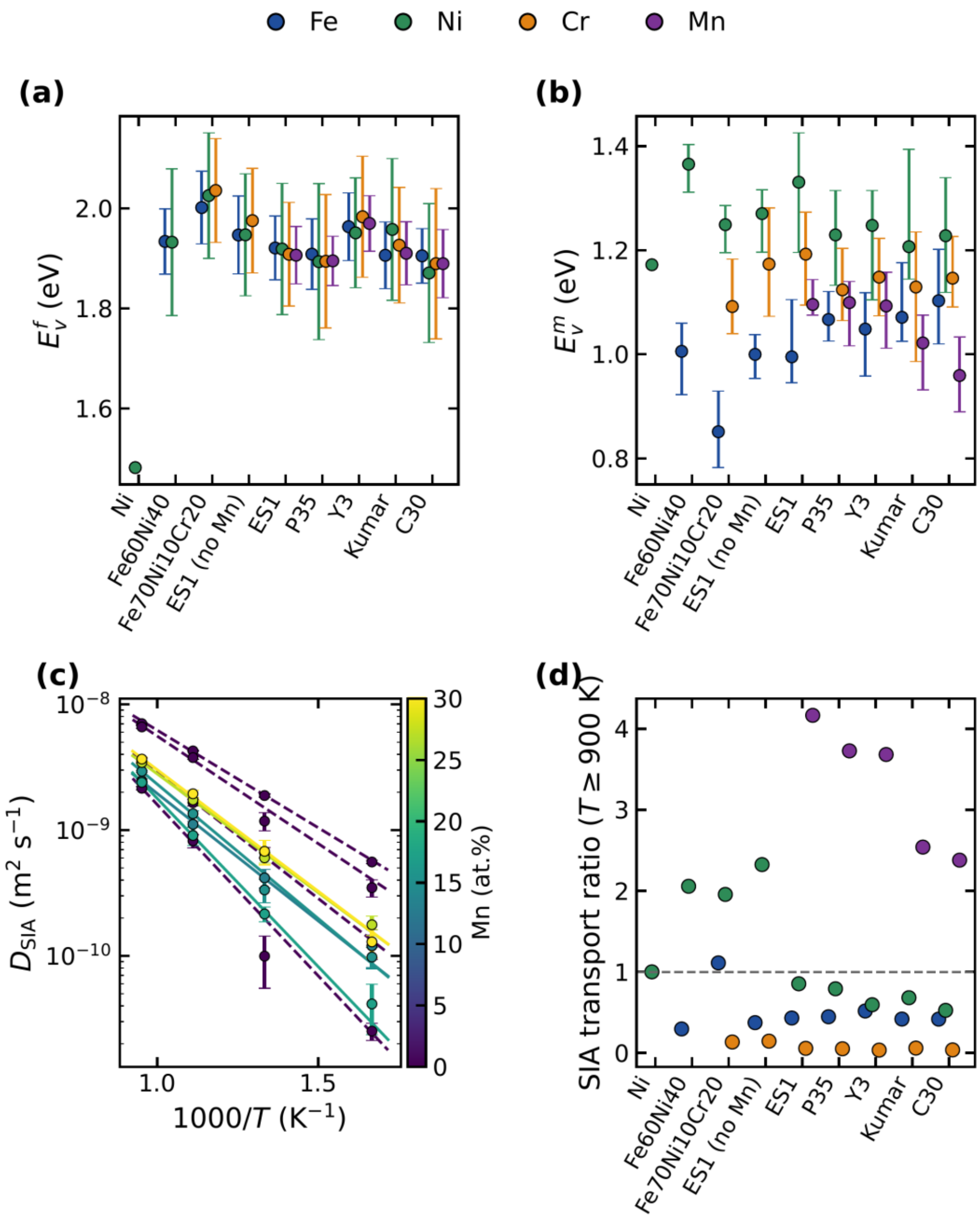

***Fig. 3.*** *Point defects. (a) Vacancy formation energy by the element removed. (b) Vacancy migration barrier by the element that jumps (median and interquartile range). (c) Interstitial diffusion coefficient against inverse temperature, with Arrhenius fits. (d) Interstitial transport ratio of each element from the runs at 900 and 1050 K; the dashed line marks a ratio of one.*

## 3.7 Frenkel pair production

Fig. 4(a) shows the mean number of Frenkel pairs against time in the 40 keV cascades. In every alloy the number rises to a peak at 0.8–1.2 ps and falls by two orders of magnitude over the next 10 ps. At 40 keV the peak is about 7000 in Ni, 4500 in $Fe_{60}Ni_{40}$ and 5200–6400 in the other alloys. The number of surviving Frenkel pairs is fixed after about 10 ps. Only 1.1–1.8% of the Frenkel pairs present at the peak survive (Fig. 4b). The surviving fraction is highest in Ni at 40 keV (1.7%) and does not depend on Mn: ES1 and ES1noMn give 1.3 and 1.2% at 40 keV and 1.3 and 1.4% at 20 keV.

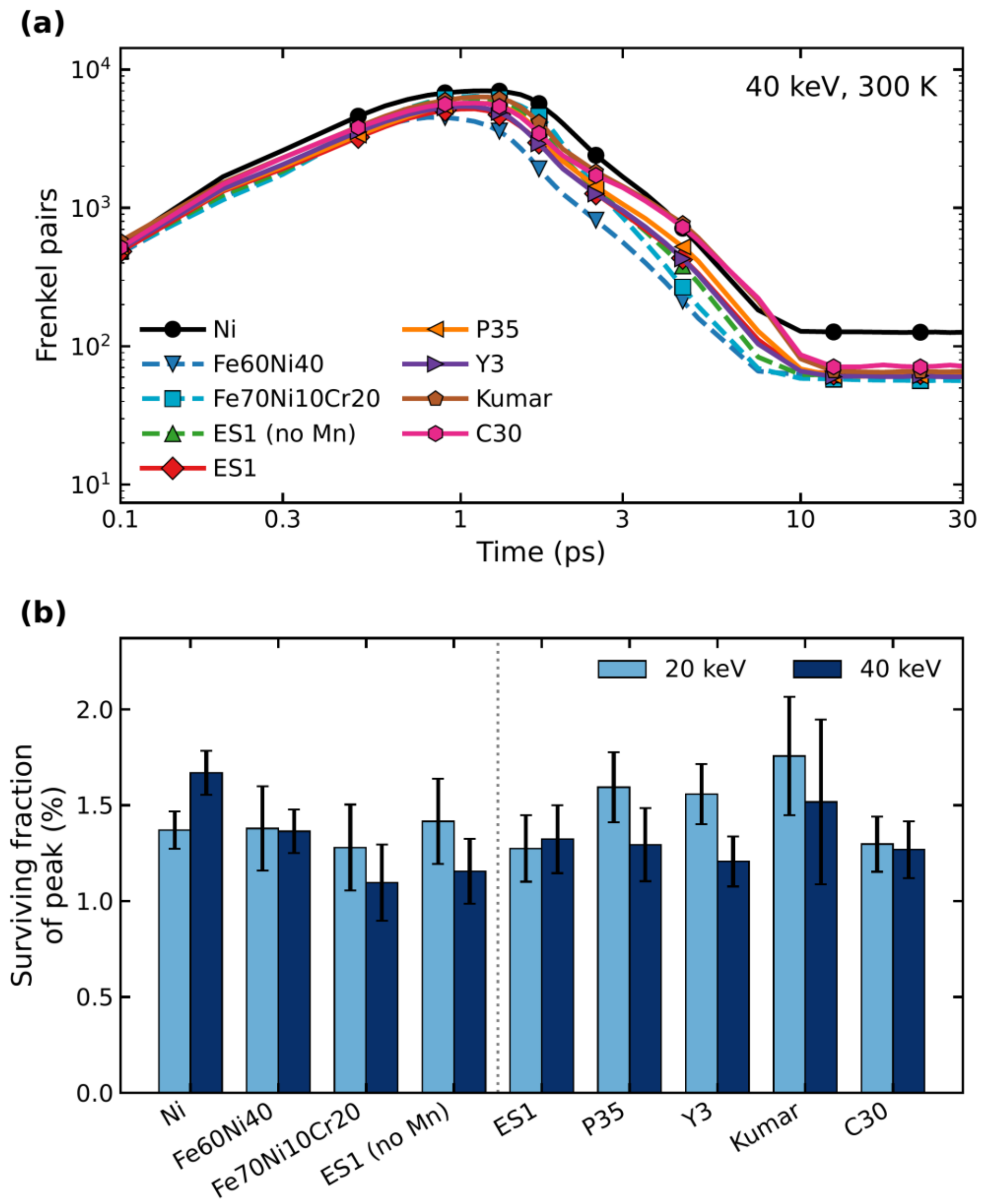

***Fig. 4.*** *Frenkel pairs against time at 300 K. (a) Mean number of Frenkel pairs in the 40 keV cascades, from Wigner–Seitz analysis. Solid lines: Ni and the Mn-containing alloys; dashed lines: Mn-free alloys. (b) Fraction of the peak number that survives at 30 ps, at 20 and 40 keV (mean and standard error over cascades).*

The surviving Frenkel pairs are given in Table 7 and Fig. 5. In the concentrated alloys their number is almost independent of composition: 6–9 at 5 keV, 12–17 at 10 keV, 27–32 at 20 keV and 56–72 at 40 keV. Ni produces more defects at every energy, and the difference grows with energy, to 127 Frenkel pairs at 40 keV, twice the alloy value. The ES1 minus ES1noMn differences are −0.8±1.1, +3.6±1.2, −1.1±1.9 and +1.3±7.5 Frenkel pairs at 5, 10, 20 and 40 keV. Mn does not change the number of surviving defects in a consistent way.

The survival efficiency ξ is 0.24–0.41 in the alloys at all energies (Fig. 5b). It lies on or above the arc-dpa curve for Ni, and it is lower in the Mn-containing alloys (0.30–0.33 at 40 keV) than in the Mn-free alloys (0.36–0.39). This difference comes from $N_{\mathrm{NRT}}$: the alloys give the same number of Frenkel pairs, but the Mn-containing alloys have lower $E_d^{\mathrm{eff}}$ and therefore a larger $N_{\mathrm{NRT}}$. In Ni, ξ rises from 0.35 at 5 and 10 keV to 0.63 at 40 keV. This is where large interstitial clusters appear in Ni (Section 3.8).

***Table 7.*** *Cascades at 40 keV and 300 K (15 per alloy): surviving Frenkel pairs (mean ± standard error), survival efficiency $\xi = N_{FP}/N_{NRT}$, fraction of interstitials ($f_{cl}^I$) and vacancies ($f_{cl}^V$) in clusters, mean size of the largest interstitial cluster, faulted fraction of the interstitials in clusters of four or more, mean DXA dislocation length per cascade, and number of cascades with defects near the cell faces.*

| Alloy | $N_{FP}$ | ξ | $f_{cl}^I$ | Largest | $f_{cl}^V$ | Faulted | DXA (Å) | Near face |
|---|---|---|---|---|---|---|---|---|
| Ni | 126.5±12.5 | 0.63 | 0.82±0.03 | 56.3 | 0.62 | 0.90±0.04 | 255 | 6 |
| Fe60Ni40 | 61.3±5.0 | 0.36 | 0.67±0.04 | 22.4 | 0.64 | 0.88±0.04 | 126 | 6 |
| Fe70Ni10Cr20 | 56.3±3.3 | 0.39 | 0.71±0.03 | 18.3 | 0.68 | 0.90±0.03 | 184 | 4 |
| ES1noMn | 61.7±5.1 | 0.38 | 0.69±0.04 | 20.5 | 0.67 | 0.89±0.05 | 179 | 2 |
| ES1 | 63.1±5.4 | 0.33 | 0.54±0.04 | 13.4 | 0.58 | 0.76±0.08 | 101 | 8 |
| P35 | 60.8±4.5 | 0.30 | 0.55±0.05 | 14.9 | 0.62 | 0.72±0.05 | 100 | 4 |
| Y3 | 60.3±3.8 | 0.31 | 0.57±0.04 | 16.4 | 0.68 | 0.70±0.08 | 105 | 4 |
| Kumar | 65.9±6.6 | 0.30 | 0.57±0.04 | 16.4 | 0.71 | 0.71±0.12 | 140 | 4 |
| C30 | 71.7±8.1 | 0.32 | 0.58±0.05 | 20.9 | 0.71 | 0.82±0.10 | 165 | 4 |

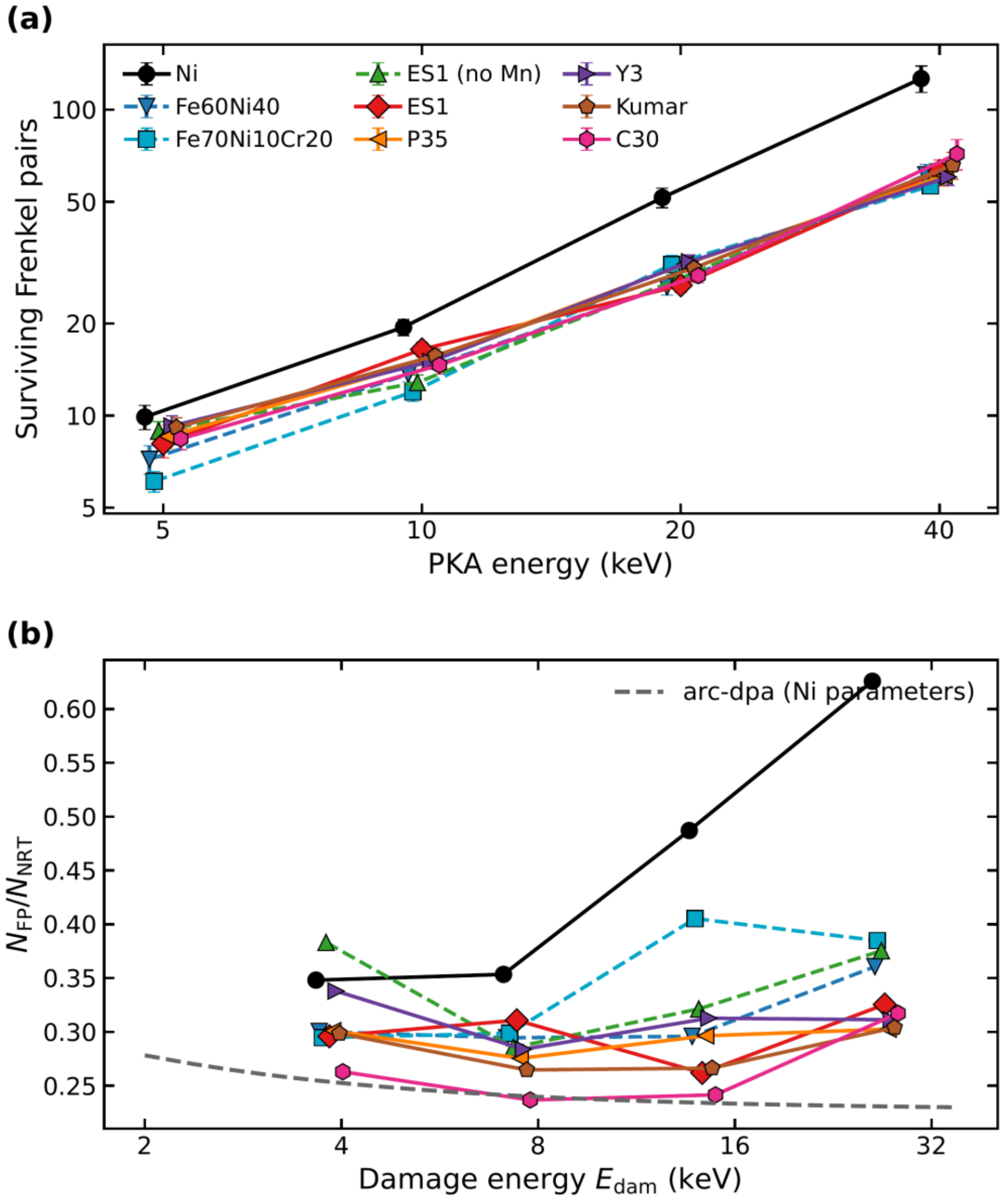

***Fig. 5.*** *Surviving Frenkel pairs at 300 K. (a) Number of Frenkel pairs against PKA energy (mean and standard error; points are shifted slightly along the energy axis for clarity). (b) Survival efficiency* $\xi = N_{FP}/N_{NRT}$ *against damage energy, with the arc-dpa curve for Ni [47].*

## 3.8 Defect clusters

The fraction of interstitials in clusters and the largest interstitial cluster are shown in Fig. 6. Both increase with energy in every alloy. Ni has the most clustered interstitials (0.82 at 40 keV) and the largest clusters, 56 interstitials on average at 40 keV. Among the concentrated alloys there are two groups. In the Mn-free alloys 0.67–0.71 of the interstitials are in clusters at 40 keV, and the largest cluster has 18–22 interstitials. In the five Mn-containing alloys the fraction is 0.54–0.58 and the largest cluster has 13–21 interstitials. The same grouping is seen at 10 and 20 keV, where the Mn-free alloys have 0.49–0.57 and 0.57–0.66 and the Mn-containing alloys 0.35–0.43 and 0.48–0.53. ES1 has the lowest fraction among all the alloys at 40 keV. Compared with ES1noMn, the fraction of clustered interstitials in ES1 is lower by 0.07±0.08, 0.12±0.05, 0.08±0.05 and 0.15±0.05 at 5, 10, 20 and 40 keV. The sign is the same at every energy.

Vacancy clustering does not show this grouping. At 40 keV, 0.58–0.71 of the vacancies are in clusters, and the Mn-containing Kumar and C30 have the highest values (Table 7). The effect of Mn is therefore specific to the interstitials.

The cascades that left defects near the cell faces (119 of 1015) do not change these conclusions. Without them, the clustered fraction at 40 keV is 0.55 in ES1, 0.73 in $Fe_{60}Ni_{40}$, 0.74 in $Fe_{70}Ni_{10}Cr_{20}$ and 0.89 in Ni.

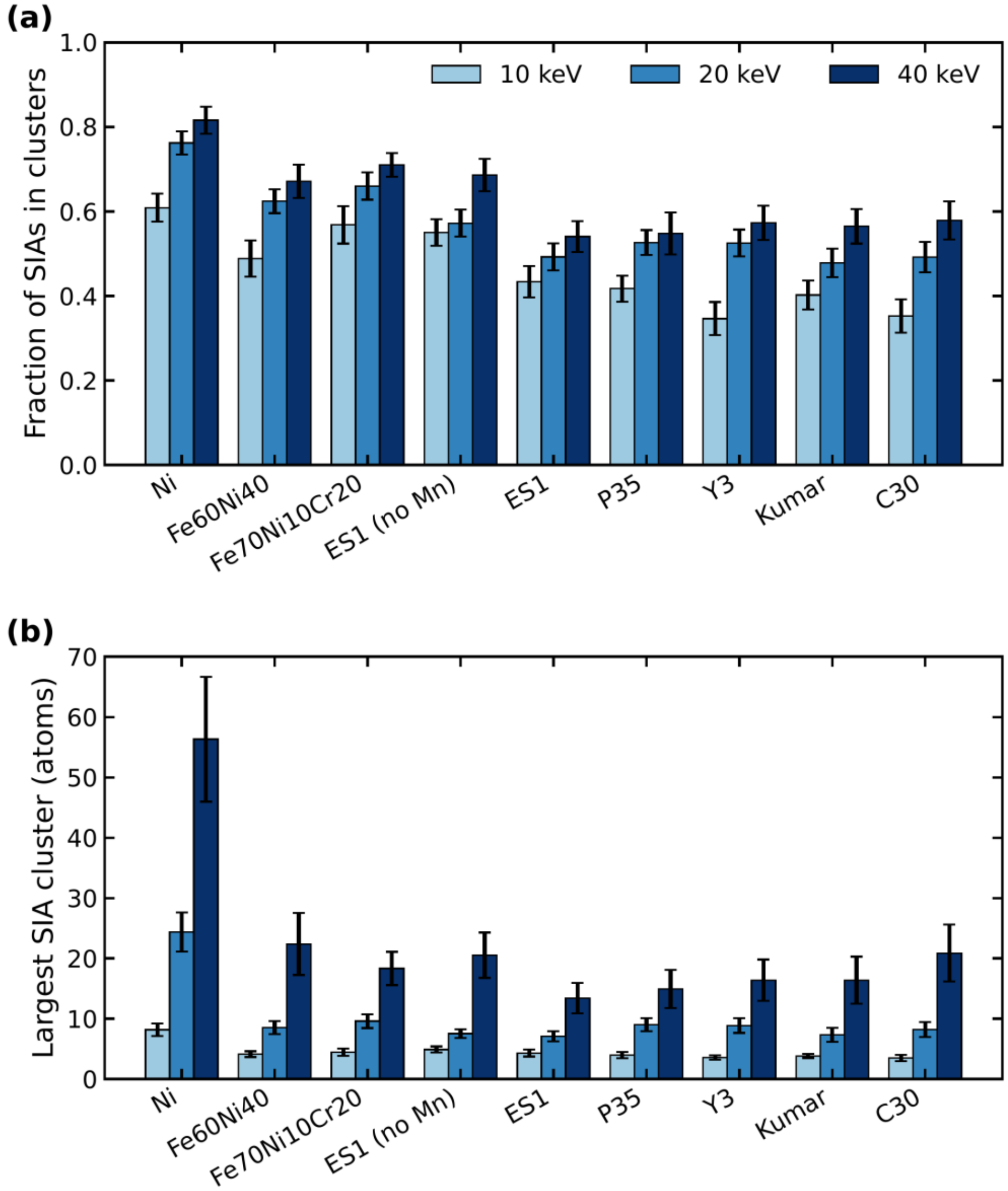

***Fig. 6.*** *Interstitial clusters at 300 K. (a) Fraction of surviving interstitials in clusters of two or more. (b) Size of the largest interstitial cluster per cascade. Mean and standard error over cascades.*

Fig. 7 shows the damage left by typical 40 keV cascades in Ni, $Fe_{70}Ni_{10}Cr_{20}$, ES1noMn and ES1. In Ni most interstitials are in one large cluster bounded by Frank and Shockley segments, next to a core of vacancies. In $Fe_{70}Ni_{10}Cr_{20}$ and ES1noMn the interstitials are in several compact clusters, the largest with 21 and 20 interstitials, bounded mainly by Shockley and stair-rod segments. In ES1 more interstitials are isolated, and the largest cluster is smaller. In the close-ups, the clusters in Ni, $Fe_{70}Ni_{10}Cr_{20}$ and ES1noMn contain planar groups of hcp atoms, which mark a stacking fault, while the ES1 cluster contains only a few.

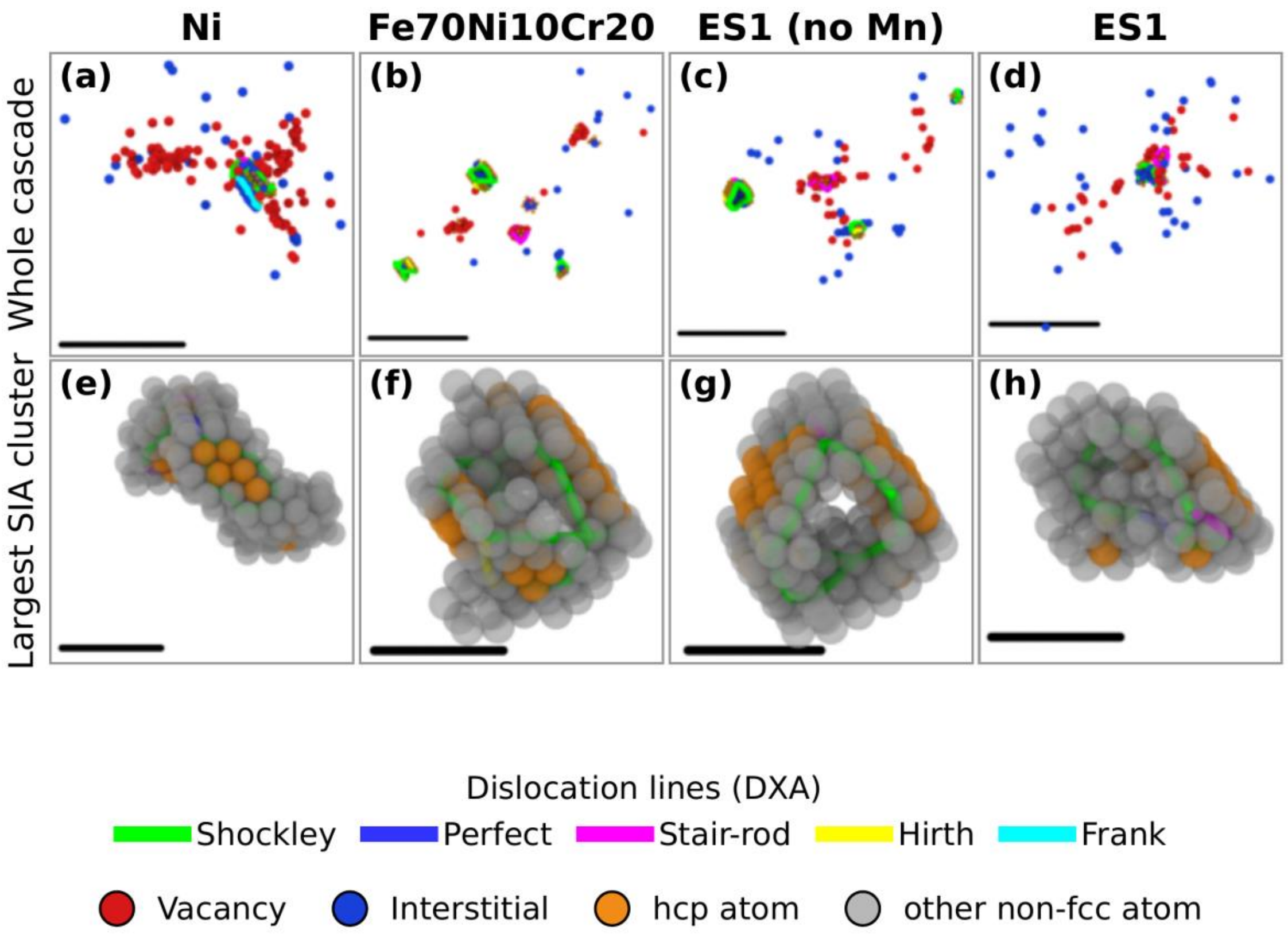

***Fig. 7.*** *Damage at 30 ps in typical 40 keV cascades at 300 K. Top row (a–d): vacancies (red), interstitials (blue), hcp atoms after relaxation (orange) and DXA dislocation lines coloured by type; the scale bar is 5 nm. Bottom row (e–h): non-fcc atoms around the largest interstitial cluster of the same cascade, with the dislocation lines inside; hcp atoms are orange and other non-fcc atoms grey. The scale bar is 1 nm. The largest clusters contain 54, 21, 20 and 16 interstitials.*

## 3.9 Faulted clusters and dislocation content

At 40 keV, 0.88–0.90 of the interstitials in clusters of four or more are in faulted clusters in Ni and the Mn-free alloys, and 0.70–0.82 in the Mn-containing alloys (Table 7). The faulted fraction does not follow the SFE (Fig. S3, Table S3). $Fe_{70}Ni_{10}Cr_{20}$ (19 mJ m$^{-2}$) and Ni (113 mJ m$^{-2}$) both give 0.90, and the Mn-containing alloys, with intermediate SFE, give the lowest values.

Large clusters are almost always faulted, so a lower faulted fraction could simply reflect the smaller clusters in the Mn-containing alloys. Table 8 and Fig. 8(a) therefore give the faulted fraction of clusters by size, with the 20 and 40 keV cascades pooled. Clusters of 13 or more interstitials are faulted in nearly every case in all alloys (0.88–1.00). For smaller clusters the two groups differ. The Mn-free concentrated alloys together have 0.58 of the clusters of 4–6 interstitials faulted (87 of 151) and the Mn-containing alloys 0.24 (51 of 210). For clusters of 7–12 the values are 0.81 (65 of 80) and 0.46 (44 of 95). Both differences are significant ($p<10^{-5}$, two-proportion test). For clusters of 4–12, ES1 gives 0.43 (26 of 61) and ES1noMn 0.56 (40 of 72). Pure Ni behaves differently: its small clusters are rarely faulted (0.13 for 4–6), while its large clusters are all faulted.

***Table 8.*** *Faulted fraction of interstitial clusters by cluster size, 20 and 40 keV cascades at 300 K pooled, with the number of faulted clusters and the total number of clusters in brackets. Last columns: mean DXA line length per 40 keV cascade by Burgers vector (Å).*

| | Cluster size (interstitials) | | | DXA line length (Å) | | | |
|---|---|---|---|---|---|---|---|
| Alloy | 4–6 | 7–12 | ≥13 | Shockley | Stair-rod | Frank | Total |
| Ni | 0.13 (7/55) | 0.55 (21/38) | 0.98 (57/58) | 131 | 52 | 47 | 255 |
| Fe60Ni40 | 0.54 (27/50) | 0.85 (22/26) | 1.00 (13/13) | 60 | 39 | 15 | 126 |
| Fe70Ni10Cr20 | 0.70 (37/53) | 0.87 (26/30) | 1.00 (24/24) | 103 | 58 | 5 | 184 |
| ES1noMn | 0.48 (23/48) | 0.71 (17/24) | 0.94 (16/17) | 108 | 48 | 8 | 179 |
| ES1 | 0.36 (15/42) | 0.58 (11/19) | 1.00 (13/13) | 44 | 41 | 7 | 101 |
| P35 | 0.19 (8/43) | 0.56 (10/18) | 0.89 (16/18) | 43 | 42 | 5 | 100 |
| Y3 | 0.24 (12/51) | 0.57 (13/23) | 0.88 (15/17) | 42 | 44 | 9 | 105 |
| Kumar | 0.17 (6/35) | 0.26 (6/23) | 1.00 (12/12) | 53 | 73 | 4 | 140 |
| C30 | 0.26 (10/39) | 0.33 (4/12) | 1.00 (19/19) | 60 | 76 | 16 | 165 |

The dislocation content found by DXA agrees with this (Table 8, Fig. 8b). Ni has the most dislocation line per cascade (255 Å), with the largest Frank component (47 Å). In the Mn-free alloys the total is 126–184 Å. In the Mn-containing alloys it is 100–165 Å, and the drop is mostly in the Shockley segments: 42–60 Å, against 60–108 Å without Mn. ES1 has 44 Å of Shockley line and ES1noMn 108 Å. The stair-rod length is similar in all the concentrated alloys (39–58 Å) and higher in Kumar and C30 (73–76 Å), the two alloys with the most Mn. Perfect $1/2\langle 110\rangle$ segments are rare in every alloy (below 14 Å per cascade). Dislocation lines were found in 13–15 of the 15 cascades in the Mn-free alloys and in 10–13 in the Mn-containing alloys.

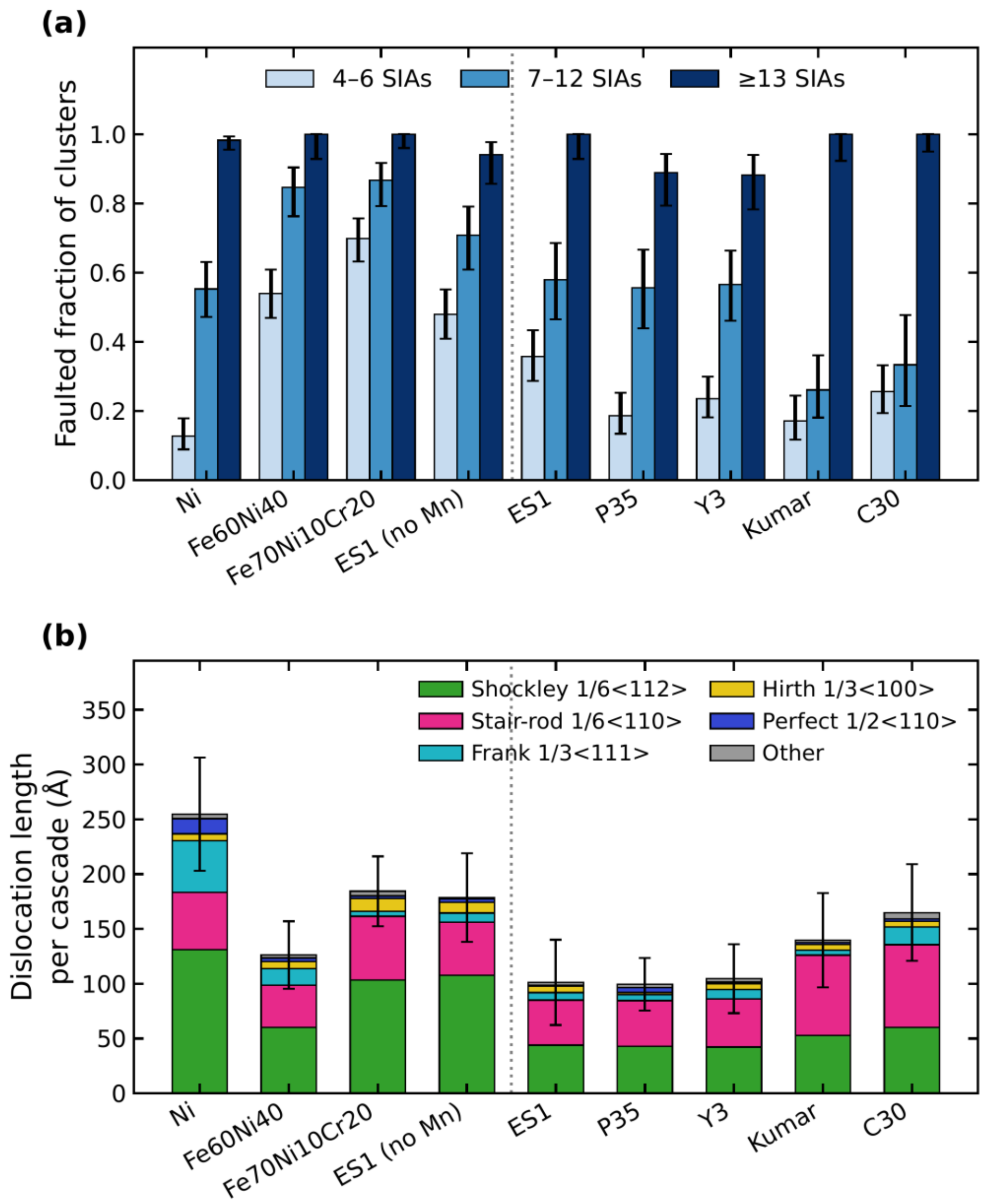

***Fig. 8.*** *Morphology of the interstitial clusters at 300 K. (a) Faulted fraction of clusters by cluster size, 20 and 40 keV pooled; error bars are 68% Wilson intervals. (b) Mean DXA dislocation line length per 40 keV cascade by Burgers vector; error bars are standard errors of the total. The dotted line separates the Mn-free and the Mn-containing alloys.*

## 3.10 Chemistry of the surviving defects

Fig. 9 gives the enrichment of each element among the surviving interstitials and vacancies, defined as the fraction of that element among the atoms of the interstitial dumbbells, or among the atoms originally at the vacant sites, divided by its concentration. The interstitials are enriched in some elements. In the Mn-free alloys, Ni is enriched by a factor 2.0–2.4 and Fe and Cr are depleted (0.35–0.45 and about 0.3), except Fe in $Fe_{70}Ni_{10}Cr_{20}$ (1.0). In the Mn-containing alloys Mn is enriched by 2.1–3.1, Ni is close to random (0.75–1.1), and Cr is depleted (0.09–0.14). The vacancies are much closer to random: all enrichment factors are between 0.79 and 1.16. There is a small depletion of Ni at vacancies (0.79–0.98) in most alloys. The interstitial enrichment follows the dumbbell formation energies and the transport ratios (Section 3.6). The chemistry of the interstitials is therefore set during the cascade, before any long-range diffusion.

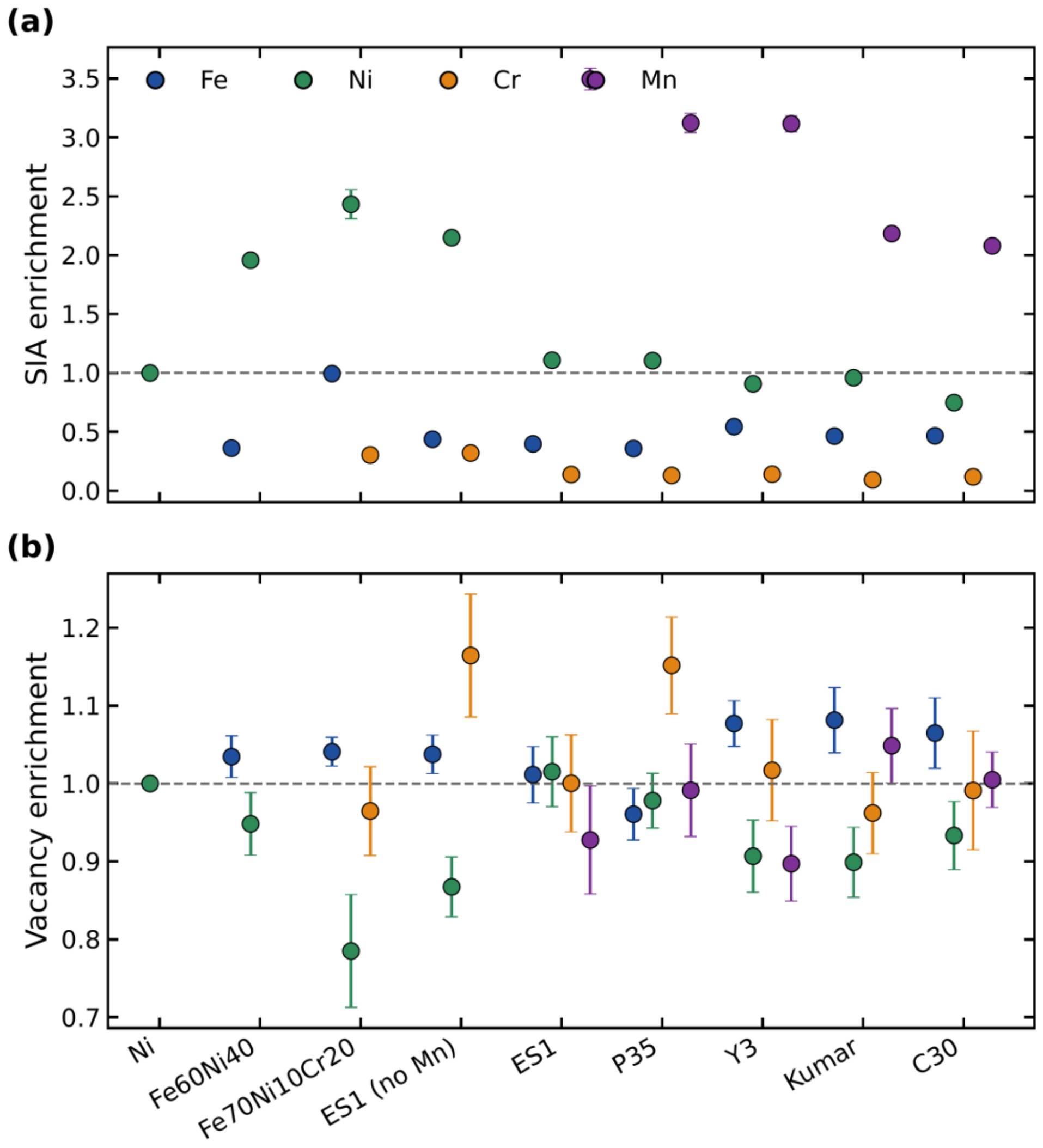

***Fig. 9.*** *Enrichment of each element among (a) the atoms of the surviving interstitial dumbbells and (b) the atoms originally at the vacant sites, relative to the alloy composition, for all cascades at 300 K. A value of one (dashed line) means no enrichment.*

## 3.11 Effect of temperature

At 823 K the cascades in $Fe_{70}Ni_{10}Cr_{20}$, ES1noMn, ES1 and Y3 left 9–21% fewer Frenkel pairs than at 300 K (Fig. 10a). The clustered fraction of interstitials changed little (Fig. 10b). It increased slightly in ES1noMn, ES1 and Y3, for example from 0.43 to 0.51 at 10 keV in ES1, and decreased in $Fe_{70}Ni_{10}Cr_{20}$ at 10 keV. The difference between ES1 and ES1noMn remained at 823 K: 0.51 against 0.61 at 10 keV and 0.57 against 0.66 at 20 keV. Over 30 ps there is little interstitial migration even at 823 K, so these numbers describe the damage at the end of the cascade and not after long-range diffusion.

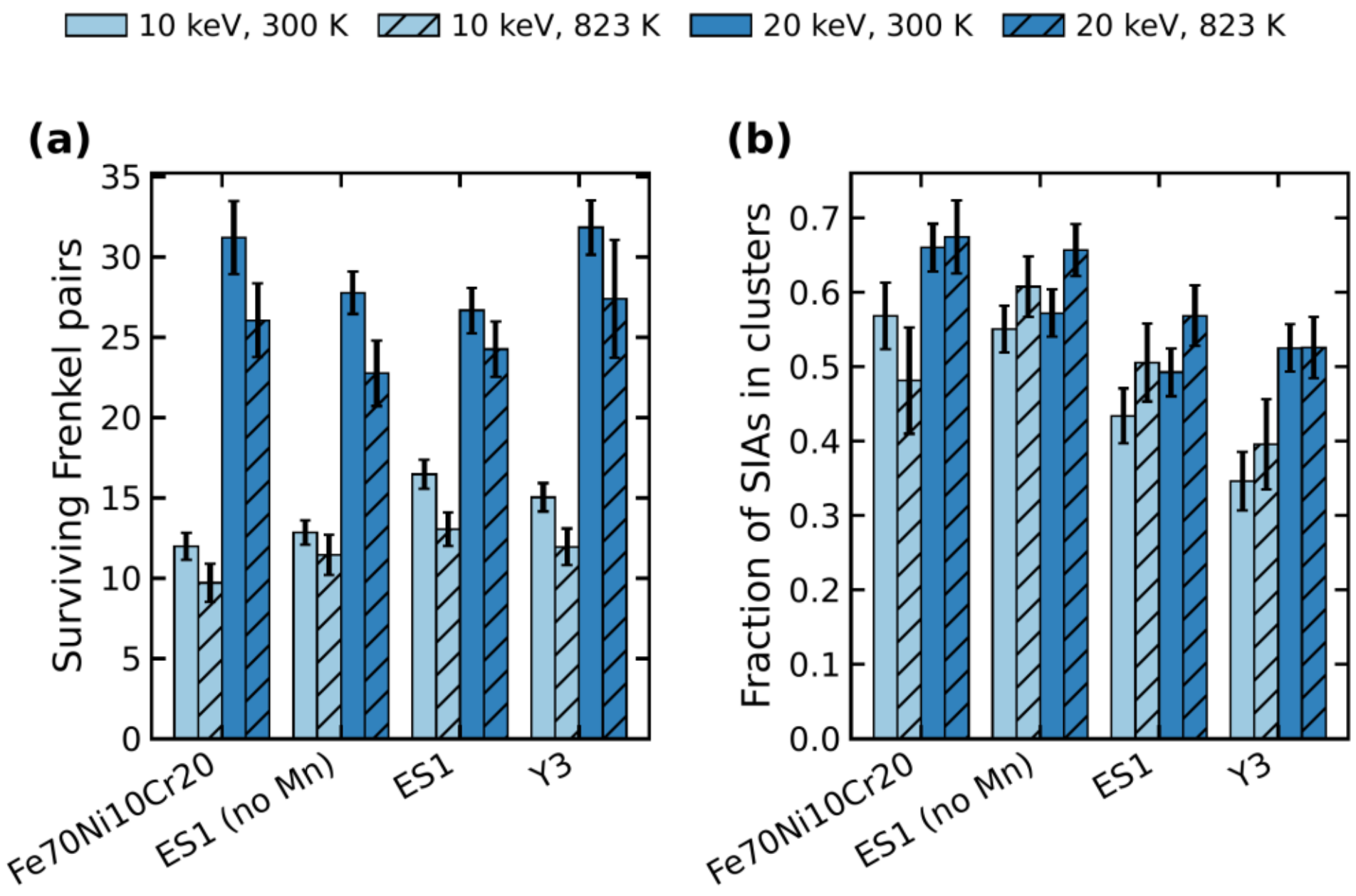

***Fig. 10.*** *Effect of temperature in $Fe_{70}Ni_{10}Cr_{20}$, ES1noMn, ES1 and Y3 at 10 and 20 keV. (a) Surviving Frenkel pairs. (b) Fraction of interstitials in clusters. Plain bars: 300 K; hatched bars: 823 K.*

# 4 Discussion

## 4.1 Number of defects

At 40 keV the concentrated alloys leave half as many Frenkel pairs as Ni and do not form the large interstitial clusters found in Ni. Granberg et al. and Lin et al. found the same difference between Ni and equiatomic alloys [3,24], and Nair et al. found more defects in Ni than in five Fe–Ni–Cr alloys [26]. Among the concentrated alloys the number of Frenkel pairs hardly depends on composition, and Mn does not change it at fixed Ni and Cr. Deluigi et al. also found the same Frenkel-pair production in an equiatomic alloy and in its average-atom counterpart [25].

The TDE differs by up to 26 eV between the alloys (Table 3), but the number of Frenkel pairs does not follow it. A lower $E_d^{\mathrm{eff}}$ increases $N_{\mathrm{NRT}}$, so ξ is lower in the Mn-containing alloys while $N_{\mathrm{FP}}$ stays the same. About 98–99% of the Frenkel pairs present at the peak recombine (Fig. 4), so the number that survives is set by the recombination stage and not by the threshold events. A dose calculated with the $E_d$ of each alloy would therefore give differences between the alloys that are not present in the cascades.

## 4.2 Direct formation of interstitial clusters

Dartois et al. concluded from in situ irradiation at 550 ˚C that interstitial clusters form directly in cascades in $Fe_{60}Ni_{40}$ and Fe–10Ni–20Cr but not in ES1 [12]. In our cascades clusters form in all three alloys, but fewer of them in ES1. At 40 keV, 0.67 and 0.71 of the interstitials are in clusters in $Fe_{60}Ni_{40}$ and $Fe_{70}Ni_{10}Cr_{20}$ and 0.54 in ES1, and the largest clusters have 22, 18 and 13 interstitials. ES1noMn behaves like the Mn-free alloys, so the lower clustering in ES1 comes from Mn and not from its Ni content. Each cascade in ES1 therefore produces fewer and smaller loop nuclei. At 550 ˚C small clusters are also more likely to dissolve or to be absorbed by existing loops before they become visible. Both effects agree with the observation that new loops kept appearing in $Fe_{60}Ni_{40}$ but not in ES1 [12]. The difference between ES1 and ES1noMn is still present at 823 K (Section 3.11). Loop growth and coarsening, also measured by Dartois et al., take place after the cascade and are not treated here.

## 4.3 Faulted clusters, Mn and stacking fault energy

The faulted fraction does not follow the SFE (Fig. S3). $Fe_{70}Ni_{10}Cr_{20}$, with the lowest SFE, and Ni, with the highest, give the same value. At a given cluster size the faulted fraction depends on Mn. Clusters of 4–12 interstitials are faulted about half as often in the Mn-containing alloys as in the Mn-free alloys. The ES1/ES1noMn pair shows the same change, although Mn raises the SFE by only 4.8 mJ $m^{-2}$ in this pair. Clusters of 13 or more interstitials are faulted in all alloys. The DXA results show the same thing: the Mn-containing alloys have less Shockley line.

Nair et al. found that the SFE given by different Ni potentials changes the size of the largest clusters [26]. Their comparison was between potentials for one element. Ours is between compositions with one potential, and here the composition effect does not go through the SFE. With EAM-21, Frank loops of 2 nm and larger in Y3 unfaulted less often than in $Fe_{70}Ni_{10}Cr_{20}$ when cut by edge dislocations, although Y3 has the higher SFE [22]. This was related to the larger local fluctuations of the SFE and the higher friction stress in Y3. In both cases the average SFE does not decide the loop type in these alloys.

Parkin et al. found more faulted loops in $Cr_{15}Fe_{35}Mn_{15}Ni_{35}$ (P35) than in $Cr_{18}Fe_{27}Mn_{27}Ni_{28}$ (Kumar) and related this to the lower measured SFE of P35 [10]. We find the same order. For clusters of 7–12 interstitials P35 gives 0.56 and Kumar 0.26; with all sizes at 20 and 40 keV pooled, P35 gives 0.63 and Kumar 0.58. In our calculations the SFE of P35 is the higher of the two, so the difference does not come from the SFE. It follows the Mn content, which is 15% in P35 and 27% in Kumar. The SFE may still play a role in the experiments, but the simulations show that Mn changes the type of small interstitial clusters on its own.

The interstitials in the Mn-containing alloys are rich in Mn (Fig. 9), Mn dumbbells have the lowest formation energy (Table 4), and Mn is the element moved most by interstitials (Table 6). A possible explanation is that Mn-rich groups of interstitials stay as three-dimensional groups of dumbbells instead of forming a faulted platelet, and only form a faulted loop at larger size. The higher mobility of Mn-rich interstitials would also let more of them leave the cascade region during the spike, which would explain the lower fraction in clusters and the smaller clusters.

## 4.4 Chemistry of the defects and segregation

The interstitials left by the cascades are already enriched or depleted before any diffusion. In the Mn-free alloys they are rich in Ni and poor in Cr, and in the Mn-containing alloys rich in Mn and poor in Cr. The vacancies are close to random. Gao et al. measured Ni enrichment and Cr, Fe and Mn depletion at loops in Y3 at 823 K [9], and Kumar et al. found weak segregation at grain boundaries in $Cr_{18}Fe_{27}Mn_{27}Ni_{28}$ [8]. The Cr depletion of the interstitials agrees with the Cr depletion measured at loops, and the Ni enrichment in the Mn-free alloys with the measured Ni enrichment. In the Mn-containing alloys the interstitials carry Mn, so the Mn depletion measured at loops at 823 K [9] must come from the vacancy flux, which acts over much longer times than a cascade. Mn jumps have some of the lowest vacancy migration barriers and Ni jumps the highest (Table 5). A vacancy flux to sinks would therefore take Mn away from them and leave Ni behind, as in the measured profiles.

All alloys were random solid solutions. Short-range order has been seen after neutron irradiation of C30 [11], first-principles Monte Carlo simulations predict Mn–Ni ordering in these alloys [20], and short-range order changes primary damage in CrFeNi [28]. Since Mn controls interstitial clustering here, Mn–Ni order may change the size of the effect. Each cascade was run in a perfect crystal for 30 ps. Cascade overlap, which controls loop growth at higher dose [31], and the later diffusion of the defects were not included. The faulted fraction depends on the thresholds used to define a faulted cluster; the same thresholds were used for all alloys, and the Mn effect appears both in the hcp atoms from PTM and in the Shockley line length from DXA. So future work will look into improving the potential to reproduce the Short-range order effect calculated in [20].

# 5 Conclusions

We compared five Co-free Fe–Ni–Cr–Mn HEAs (ES1, P35, Y3, Kumar and C30) with pure Ni, $Fe_{60}Ni_{40}$, the $Fe_{70}Ni_{10}Cr_{20}$ model alloy of 316L and the Mn-free version of ES1 (ES1noMn), using one stiffened EAM potential. The main results are as follows.

1. The SFE follows the Ni content, from 19 mJ m$^{-2}$ in $Fe_{70}Ni_{10}Cr_{20}$ to 44 mJ m$^{-2}$ in $Fe_{60}Ni_{40}$, with the HEAs in between (31–41 mJ m$^{-2}$) and Ni at 113 mJ m$^{-2}$. At fixed Ni and Cr, 13% Mn raises it by 4.8±1.5 mJ m$^{-2}$.

2. $E_d^{\mathrm{eff}}$ is 48–56 eV in the HEAs, 64–74 eV in $Fe_{60}Ni_{40}$, $Fe_{70}Ni_{10}Cr_{20}$ and ES1noMn, and 54 eV in Ni. At fixed Ni and Cr, Mn lowers the TDE of Ni and Cr recoils by about 6 eV.

3. The vacancy formation energy is 1.89–2.01 eV in all the concentrated alloys and 1.48 eV in Ni. Ni has the highest vacancy migration barrier in every alloy. In the HEAs Mn dumbbells have the lowest formation energy and interstitials move mainly Mn; in $Fe_{60}Ni_{40}$, $Fe_{70}Ni_{10}Cr_{20}$ and ES1noMn they move mainly Ni. Cr is hardly moved by interstitials in any alloy.

4. The HEAs and the Mn-free concentrated alloys leave the same number of Frenkel pairs, 56–72 at 40 keV, which is half the number in Ni. Mn has no consistent effect on this number.

5. Mn reduces the clustering of interstitials. At 40 keV, 0.54–0.58 of the interstitials are in clusters in the HEAs, 0.67–0.71 in $Fe_{60}Ni_{40}$, $Fe_{70}Ni_{10}Cr_{20}$ and ES1noMn, and 0.82 in Ni. The largest clusters have 13–21 interstitials in the HEAs, 18–22 in the Mn-free alloys and 56 in Ni. The ES1/ES1noMn difference has the same sign at every energy.

6. At the same cluster size, clusters of 4–12 interstitials are faulted about half as often in the HEAs as in the Mn-free concentrated alloys (0.24 against 0.58 for 4–6 interstitials and 0.46 against 0.81 for 7–12). This does not follow the SFE. Clusters of 13 or more interstitials are faulted in all alloys. In Ni small clusters are rarely faulted and large ones always are.

7. The surviving interstitials are enriched in Mn in the HEAs and in Ni in the Mn-free alloys, and depleted in Cr in all alloys, before any long-range diffusion. The vacancies are close to random.

The results support direct cluster formation in cascades in $Fe_{60}Ni_{40}$ and $Fe_{70}Ni_{10}Cr_{20}$ and less of it in ES1, as proposed by Dartois et al. They also show that in the HEAs Mn changes the type of small interstitial clusters at the same SFE. Cascade overlap at higher dose and the later evolution of the faulted and unfaulted clusters will be studied next.

## Supplementary information

Figs. S1–S3 and Tables S1–S3 are given in the Electronic Supplementary Material.

## Acknowledgements

AD received support from the French ANR-PRCE-HERIA project (ANR-19-CE08-0012-01) to develop the EAM potential. Some of the alloys studied were developed with co-funding from EDF (2018, master thesis of S. Boulila); AF acknowledges this support. GB acknowledges funding from the Euratom research and training programme 2014–2018 under grant agreement No. 755269 (GEMMA project). AF acknowledges funding from INNUMAT (Grant Agreement No. 101061241). The work of DNM was funded by the Engineering and Physical Sciences Research Council (EPSRC) Fusion Grant 2022/27 [grant number EP/W006839/1]. DNM and AD acknowledge EUROfusion for access to the PITAGORA high-performance computing facilities at CINECA, Bologna, Italy.

## Declarations

**Funding** The funding sources are listed in the Acknowledgements.

**Conflict of interest** The authors declare that they have no conflict of interest.

**Data availability** The stiffened potential, the LAMMPS input files, the analysis scripts and the data behind the figures are available from the corresponding authors on reasonable request.

**Author contributions** AD: conceptualisation, simulations, analysis, writing of the original draft. GB, GA, CD, GM and AF: review and editing. AD and GB: development of the EAM-21 potential, review and editing. DNM: supervision, conceptualisation, review and editing.